\documentstyle[12pt]{article}
\newcommand{\bra}{\langle}
\newcommand{\ket}{\rangle}
\newcommand{\R}{\mbox{\boldmath $ R $}}
\newcommand{\C}{\mbox{\boldmath $ C $}}
\newcommand{\Z}{\mbox{\boldmath $ Z $}}
\newcommand{\Rplus}{\mbox{\boldmath $ R $}_{\ge 0}}

\newcommand{\Kernel}{\mbox{Ker} \,}
\newcommand{\Image}{\mbox{Im} \,}
\newcommand{\End}{\mbox{End} \,}
\newcommand{\id}{\mbox{id}}
\newcommand{\Lie}{\mbox{\bf g}}
\newcommand{\LieSO}{\mbox{\bf so}}
\newcommand{\LieU}{\mbox{\bf u}}

\def\dfrac#1#2{\displaystyle\frac{#1}{#2}}

\newcommand{\isoarrow}
{\mathop{\hbox to 5mm{\rightarrowfill}}
\limits^{\scriptstyle \sim}}

\makeatletter
\@addtoreset{equation}{section}
\makeatother
\renewcommand{\theequation}{\arabic{section}.\arabic{equation}}

\begin{document} 
\baselineskip 7mm 
\begin{flushright}
hep-th/0006150
\\
revised on 31 August 2000 and 15 January 2001
\end{flushright}
\vspace*{8mm}
\begin{center}
{\bf \Large 
Path Integrals on Riemannian Manifolds with Symmetry
and Induced Gauge Structure}\footnote{%
The title has been changed from the first manuscript,
`Path Integral Method with Symmetry'.}
\vspace{4mm} \\
Shogo Tanimura

{\it Department of Engineering Physics and Mechanics \\
Kyoto University, Kyoto 606-8501, Japan \\
E-mail: tanimura@kues.kyoto-u.ac.jp
}
\vspace{24mm} \\
Abstract

\begin{minipage}[t]{130mm}
\baselineskip 7mm 
We formulate path integrals on any Riemannian manifold
which admits the action of a compact Lie group by isometric transformations.
We consider a path integral on a Riemannian manifold $ M $ on which
a Lie group $ G $ acts isometrically.
Then we show that the path integral on $ M $ is reduced to a family of
path integrals on a quotient space $ Q=M/G $
and that the reduced path integrals are completely classified 
by irreducible unitary representations of $ G $.
It is not necessary to assume that
the action of $ G $ on $ M $ is either free or transitive.
Hence our formulation is applicable to a wide class of manifolds, 
which includes inhomogeneous spaces, and
it covers all the inequivalent quantizations.
To describe the path integral on inhomogeneous space,
stratification geometry,
which is a generalization of the concept of principal fiber bundle,
is necessarily introduced.
Using it we show that
the path integral is expressed as a product of three factors;
the rotational energy amplitude, 
the vibrational energy amplitude,
and the holonomy factor.
When a singular point arises in $ Q $,
we determine the boundary condition of the path integral kernel
for a path which runs through the singularity.
\end{minipage}
\end{center}
%

\newpage
\baselineskip 6mm 
\section{Introduction}
Symmetry has always played an important role
in the theory of dynamical systems.
In Hamiltonian mechanics, if a dynamical system admits
large enough symmetry characterized by an Abelian group,
the system is reduced to a dynamical system
defined on a torus and becomes completely integrable \cite{Arnold}.
Even when the system has non-Abelian symmetry,
it is well-known \cite{Marsden-Weinstein}
that the system is reduced to a system with a smaller number 
of degrees of freedom.
In quantum mechanics, needless to say, symmetry plays a decisive
role in the spectrum analysis of Hamiltonian operators.
The Hilbert space of a quantum system is decomposed into
a series of subspaces following decomposition of the unitary representation
of the symmetry group and each subspace provides a reduced dynamical system.
On the other hand, the path integral is a useful formulation to study
global and geometric aspects of quantum mechanical systems,
and hence it is natural to examine how symmetry helps
the analysis of quantum systems in the scheme of path integral.

Our investigation of quantum systems with symmetry
is originally motivated by the study of molecular mechanics.
A molecular system is 
a quantum mechanical system which consists of several nuclei
and electrons. 
It has both translational and rotational symmetries.
The translational invariance enables us to 
separate the center-of-mass motion 
from the relative internal motion of the particles
by introducing the center-of-mass coordinate.
Thus the relative motion is described by a dynamical system
with fewer degrees of freedom.
Moreover, we may expect that the rotational invariance enables us
to divide the degrees of freedom of the relative motion
further into two components;
the rigid rotational motion of the whole molecule and 
the nonrigid vibrational motion.
But the separation of rotational and vibrational motions is
not so trivial 
as the separation of center-of-mass and relative motions.
Actually it has been proved by Guichardet \cite{Guichardet} that 
there is no coordinate system to separate 
the rotational motion from the vibrational motion.
A more sophisticated method is requested to describe these motions
and in fact differential geometry of fiber bundles and connections
provides a useful language to describe them
as shown by Iwai \cite{Iwai1,Iwai3}.
He formulated Hamiltonian mechanics and Schr{\"o}dinger equations
of reduced systems by using the theory of principal fiber bundle.

In the words of differential geometry,
the original system has a Riemannian manifold $ M $ as a configuration space
and the symmetry is characterized by a Lie group $ G $ of
isometric transformations of $ M $.
The reduced system is defined by a quotient space $ Q=M/G $
and if the action of the group $ G $ on $ M $ is free,
then the canonical projection map $ p : M \to Q $ 
naturally defines a principal fiber bundle 
and the Riemannian metric of $ M $ induces 
a connection by demanding that
the horizontal subspace is orthogonal to the fiber.
This kind of geometric setting has been examined in detail
by Iwai and Montgomery \cite{Iwai1,Montgomery}.

In the context of molecular mechanics, 
$ M $ is 
the configuration space of the molecule in the three dimensional space
and $ G $ is the group of rigid rotations
and therein 
it is natural to call the quotient space $ Q = M/G $ a shape space.
In this context, the induced connection represents nonholonomic constraint
imposed on the molecule
by conservation of angular momentum.
Non-separatability of the rotational and vibrational motions of the molecule
is a consequence of the nonvanishing curvature of the connection.
Thus the system is neatly described in terms of geometry of fiber bundle.

However, in most of the interesting models for physical application,
the action of the group $ G $ is not necessarily free 
but there are some points in $ M $
which admit nontrivial isotropy groups.
{}For example, in molecular mechanics,
the collinear configuration in which all the particles lie on a line 
is invariant under rotations around the molecule axis.
At this point the configuration space $ M $ 
fails to become a principal fiber bundle
and therefore it brings
a singular point to the shape space $ Q = M/G $,
where the ordinary local differential structure is broken.
To include such a singular case in consideration,
Davis \cite{Davis paper} introduced stratification structure
and showed that 
the set of spaces $ (M,Q) $ can be regarded 
as a collection of smooth fiber bundles.
In our previous paper \cite{TI},
by extending the concept of connection 
to be defined on a stratified manifold,
we provided a geometric setting 
to describe the reduced quantum system 
on the singular quotient space $ Q $.
Although Wren \cite{Wren} also showed that 
quantum systems on singular quotient spaces
can be described by the $ C^* $ algebra theory,
our geometric method is more concrete and intuitive than his algebraic method.

The study of path integrals on manifolds has a long expanding history,
which is impossible to review in this paper.
We quote only a few works here;
Schulman \cite{Schulman} studied the path integral of spin,
which is the path integral on $ SO(3) = SU(2) / \Z_2 $,
and noticed the existence of inequivalent quantizations.
Laidlaw and DeWitt \cite{Laidlaw} showed that
inequivalent quantizations are classified by unitary representations of
the homotopy group of the configuration space.
Mackey \cite{Mackey} elucidated classification of inequivalent quantizations
on homogeneous spaces from the point of view of representation theory.
When Ohnuki and Kitakado \cite{OK} wrote down representations of
quantization algebra of spheres concretely,
they noticed existence of inequivalent representations
and emergence of the gauge potential in the representation of momentum operators.
Although 
Marinov and Terentyev \cite{MT}
have been studying path integrals on homogeneous space,
they did not pay much attention to inequivalent quantizations.
Landsman, Linden \cite{Landsman-Linden} and we \cite{TT1,TT2}
studied path integrals on a homogeneous space $ Q = G/H $,
where $ H $ is a subgroup of the Lie group $ G $.
In this case, the projection map $ p : G \to G/H $ defines
a principal fiber bundle with a structure group $ H $.
They constructed the path integrals on $ Q $ by reducing 
the path integral on $ G $.
They also showed that the reduced path integrals are classified by
irreducible representations of the group $ H $
and that the path integral on $ Q $ naturally contains the holonomy
of the induced connection.
A recent work on the path integral on sphere
by Ikemori et. al. \cite{Ikemori} is also noticeable
for their fine use of spinor structure of the sphere.

The purpose of this paper is 
to make the reduction method of path integral
applicable to more general manifolds,
which are not necessarily homogeneous spaces,
or principal fiber bundles either.
Actually, our method is applicable to any Riemannian manifold $ M $ 
which admits the action of a compact Lie group $ G $ 
of isometric transformations.
{}In such a general case,
the singularities are realized as boundary points of the quotient space 
$ Q = M/G $.
As a byproduct of our reduction method of path integral,
we will obtain a boundary condition of the path integral kernel
for a path running through the singularity.
This paper is a natural extension of the previous paper 
by Tanimura and Iwai \cite{TI},
in which the Schr{\"o}dinger formalism for reduced systems was studied.

Our formalism is immediately applicable to quantum molecular dynamics,
in which methods to compute rotational and vibrational energy
spectra and to compute reaction cross sections are desired.
We also hope to apply our method to the gauge field theory.
The gauge field is a dynamical system which has gauge symmetry
and its physical degrees of freedom are described 
in terms of the quotient space of the gauge field configuration space 
modulo gauge transformations 
as discussed by many authors \cite{Rajeev,Mickelsson,LM1,LM2}.
We postpone these applications for further investigation.

This paper is organized as follows.
in Sec.II we will give a brief review of the general reduction method
to make the paper self-contained
and will introduce a time-evolution operator for the reduced system,
which is to be expressed in terms of the path integral.
In Sec.III we will construct an integral kernel
which bridges between the abstract time-evolution operator 
and the more concrete path integral.
In Sec.IV we will introduce stratification geometry,
which is needed as a proper language to describe path integrals
on inhomogeneous spaces.
In Sec.V we will use it
to write down the path integral explicitly.
In Sec.VI behavior of the kernel at a singular point will be examined.
To give a simple but nontrivial example, our formalism will be applied
to the quantum system on $ \R^2 $ with the $ SO(2) $ symmetry in Sec.VII.
Section VIII will be devoted to conclusion and discussion of the future work.

\section{Reduction of quantum system}
In this section 
we briefly review the method of reduction of quantum dynamical system
and make a step toward the definition of path integral of the reduced system.
The detailed explanation of the reduction method is given 
in a previous paper \cite{TI}.

Suppose that a quantum system $ ( {\cal H}, H ) $
admits a symmetry specified by a set $ ( G, T ) $.
Here $ {\cal H} $ is a Hilbert space with an inner product 
denoted by $ \bra \phi, \psi \ket $.
The Hamiltonian $ H $ is
a self-adjoint operator acting on $ {\cal H} $.
Let $ U({\cal H}) $ denote a group of the unitary operators
on the Hilbert space $ {\cal H} $.
A compact Lie group $ G $ is equipped with a homomorphism
$ T : G \to U({\cal H}) $, 
which is a unitary representation of $ G $ on $ {\cal H} $.
By symmetry we imply that commutativity $ T(g) H = H T(g) $ 
holds for any $ g \in G $.
The group $ G $ has a normalized invariant measure $ dg $
satisfying $ \int_G dg = 1 $.

To construct a reduced system 
we bring an irreducible unitary representation of 
$ G $ on another Hilbert space $ {\cal H}^\chi $,
that is a homomorphism
$ \rho^\chi : G \to U({\cal H}^\chi) $.
The superscript $ \chi $ labels all the inequivalent representations.
It has finite dimensions, $ d^\chi = \dim {\cal H}^\chi $.
Let a set
$ \{ e^\chi_1, \dots, e^\chi_d \} $ 
be an orthonormal basis of $ {\cal H}^\chi $.
A tensor product of operators
$ \rho^\chi(g) \otimes T(g) $ for $ g \in G $, which acts on
$ {\cal H}^\chi \otimes {\cal H} $, also provides
a unitary representation of $ G $.
The subspace of the invariant vectors of 
$ {\cal H}^\chi \otimes {\cal H} $, which is defined by
\begin{equation}
	( {\cal H}^\chi \otimes {\cal H} )^G
	:=
	\{ \psi \in {\cal H}^\chi \otimes {\cal H} 
	\, | \,
	( \rho^{\chi}(h) \otimes T(h) ) \psi = \psi,
	\, \forall h \in G \},
	\label{invariant vectors}
\end{equation}
is called the reduced Hilbert space.
Let us define a transformation 
$ S^\chi_i : {\cal H} \to {\cal H}^\chi \otimes {\cal H} $ by
\begin{equation}
	S^\chi_i \psi
	:= 
	\sqrt{d^\chi} \int_G dg \,
	( \rho^\chi(g) e^\chi_i ) \otimes (T(g) \psi),
	\qquad
	\psi \in {\cal H},
	\label{S}
\end{equation}
which then satisfies
\begin{equation}
	( \rho^{\chi}(h) \otimes T(h) ) 
	S^\chi_i = S^\chi_i,
	\quad \forall h \in G,
\end{equation}
and therefore it has an image 
$ \Image S^\chi_i \subset ( {\cal H}^\chi \otimes {\cal H} )^G $.
Actually, it can be shown that
$ \Image S^\chi_i = ( {\cal H}^\chi \otimes {\cal H} )^G $.
Using the family of transformations $ \{ S^\chi_i \}_{ (\chi,i) } $
labeled by the inequivalent representations $ \{ \rho^\chi \} $
and the basis vectors $ \{ e^\chi_i \} $,
we obtain an orthogonal decomposition of the Hilbert space
\begin{equation}
	{\cal H} 
	\cong
	\bigoplus_{\chi, i} \, \Image \, S^\chi_i
	=
	\bigoplus_{\chi, i} \, ({\cal H}^\chi \otimes {\cal H})^G.
	\label{decomposition}
\end{equation}
Since these statements have been already proven in the previous paper \cite{TI},
we omit the proof here.

Let us introduce another operator 
$ P^\chi : 
{\cal H}^\chi \otimes {\cal H} \to {\cal H}^\chi \otimes {\cal H} $,
which is defined by
\begin{equation}
	P^\chi 
	:= \int_G dg \, \rho^\chi(g) \otimes T(g).
	\label{P^chi}
\end{equation}
It can be easily seen that
\begin{eqnarray}
	&&
	(P^\chi)^\dagger = P^\chi,
	\quad
	(P^\chi)^2 =P^\chi,
	\\
	&&
	\Image P^\chi = ({\cal H}^\chi \otimes {\cal H})^G.
\end{eqnarray}
Hence $ P^\chi $ is a projection operator 
$ {\cal H}^\chi \otimes {\cal H} \to ({\cal H}^\chi \otimes {\cal H})^G $.
Then an equation
\begin{equation}
	P^\chi S^\chi_i = S^\chi_i
\end{equation}
holds.

The range of the Hamiltonian
$ H : {\cal H} \to {\cal H} $ is trivially extended to
$ (\id \otimes H) : 
{\cal H}^\chi \otimes {\cal H} \to {\cal H}^\chi \otimes {\cal H} $.
The symmetry $ T(g) H = H T(g) $ implies that
\begin{eqnarray}
	S^\chi_i H & = & ( \id \otimes H ) S^\chi_i,
	\label{invariance1}
	\\
	P^\chi ( \id \otimes H ) & = & ( \id \otimes H ) P^\chi.
	\label{invariance2}
\end{eqnarray}
The time-evolution operator $ U(t) $ on $ {\cal H} $
is defined as a unitary operator $ U(t) := e^{-iHt} $ in the usual way.
Then $ (\id \otimes U(t)) $ becomes
a time-evolution operator on $ {\cal H}^\chi \otimes {\cal H} $.
As a consequence of the invariance (\ref{invariance2}), 
$ U^\chi(t) := P^\chi ( \id \otimes U(t)) $ also provides
a unitary time-evolution operator on $ ({\cal H}^\chi \otimes {\cal H})^G $.

We conclude this short section by summarizing the above discussion;
the space of the invariant vectors $ ({\cal H}^\chi \otimes {\cal H})^G $
provides a closed dynamical system 
governed by the projected Hamiltonian 
$ H^\chi := P^\chi ( \id \otimes H ) $.
Thus we obtain decomposition of the original quantum system
$ ( {\cal H}, H ) $
into a family of quantum systems
$ \{ ( ({\cal H}^\chi \otimes {\cal H})^G, H^\chi ) \}_\chi $.
The operator $ S^\chi_i $, which reduces the original Hilbert space
to the invariant subspace,
is called the reduction operator.
The main purpose of this paper is to develop a path integral expression for
the reduced time-evolution operator
$ U^\chi(t) = P^\chi ( \id \otimes U(t)) $.

\section{Reduction of kernel}
This section is a step toward the definition of 
the reduced path integral.
Let us assume that the Hilbert space $ {\cal H} $ is 
the space of square integrable functions $ L_2(M) $ 
on a measured space $ (M, dx) $.
Assume that
the compact Lie group $ G $ acts on $ M $ by preserving the measure $ dx $.
The symmetry operation $ T(g) $ is represented on $ f \in L_2(M) $ by
\begin{equation}
	( T(g) f ) (x) := f(g^{-1}x).
	\label{T(g)}
\end{equation}
Moreover, the canonical projection map $ p : M \to Q = M/G $ induces
a measure $ dq $ of the quotient space $ Q $ as follows:
let $ \phi(q) $ is a function on $ Q $ such that
$ \phi(p(x)) $ is a measurable function on $ M $.
The measure $ dq $ of $ Q $ is then defined by
\begin{equation}
	\int_Q dq \, \phi(q) := \int_M dx \, \phi(p(x)).
	\label{dq}
\end{equation}

Suppose that the time-evolution operator $ U(t) $ 
is expressed in terms of an integral kernel 
$ K: M \times M \times \R \to \C $ as
\begin{equation}
	( U(t) f ) (x) = \int_M dy \, K(x,y;t) f(y)
\end{equation}
for any $ f(x) \in L_2(M) $.
The symmetry $ T(g) H = H T(g) $ implies a symmetry of the kernel
\begin{equation}
	K(g^{-1}x,y;t) = K(x,gy;t), \qquad g \in G,
\end{equation}
or
\begin{equation}
	K(gx,gy;t) = K(x,y;t), \qquad g \in G.
\end{equation}

On the other hand, 
a vector $ \psi \in {\cal H}^\chi \otimes L_2(M) $ 
can be identified with a measurable map
$ \psi : M \to {\cal H}^\chi \otimes \C \cong {\cal H}^\chi $.
Then with the representation (\ref{T(g)})
the tensor product operator $ \rho^\chi(g) \otimes T(g) $
acts on $ \psi $ as
\begin{equation}
	( (\rho^\chi(g) \otimes T(g)) \psi ) (x)
	=
	\rho^\chi(g) \psi (g^{-1} x),
	\qquad
	g \in G.
\end{equation}
Therefore, 
following the definition (\ref{invariant vectors}),
an invariant vector 
$ \psi \in ({\cal H}^\chi \otimes L_2(M))^G $ satisfies
\begin{equation}
	( (\rho^\chi(g) \otimes T(g)) \psi ) (x)
	=
	\rho^\chi(g) \psi (g^{-1} x)
	=
	\psi(x),
\end{equation}
or equivalently,
\begin{equation}
	\psi(gx) = \rho^\chi(g) \psi (x).
	\label{equivariant}
\end{equation}
A function $ \psi : M \to {\cal H}^\chi $
satisfying the property (\ref{equivariant}) is called an equivariant function.
Hence the reduced Hilbert space is identified with the space of 
the equivariant functions $ L_2(M; {\cal H}^\chi)^G $.
The projection operator 
$ P^\chi : L_2(M; {\cal H}^\chi) \to L_2(M; {\cal H}^\chi)^G $,
which was defined at (\ref{P^chi}) in a general form, is now given by
\begin{equation}
	(P^\chi \psi)(x)
	= \int_G dg \, \rho^\chi(g) \psi(g^{-1}x).
\end{equation}
Then the time-evolution operator of the reduced system 
is expressed as
\begin{equation}
	( P^\chi (\id \otimes U(t)) \psi ) (x) 
	= 
	\int_G dg \int_M dy \, 
	\rho^\chi(g) K(g^{-1}x,y;t) \psi(y).
\end{equation}
Therefore we define a map 
$ K^\chi : M \times M \times \R \to \mbox{End} \, {\cal H}^\chi $ by
\begin{equation}
	K^\chi(x,y;t)
	:= 
	\int_G dg \, \rho^\chi(g) K(g^{-1}x,y;t),
	\label{equivariant kernel}
\end{equation}
which we call an integral kernel of the reduced time-evolution operator
$ U^\chi(t) = P^\chi (\id \otimes U(t)) $.
It is also equivariant; namely, for any $ h \in G $, it satisfies
\begin{eqnarray}
	K^\chi(hx,y;t) & = & \rho^\chi(h) K^\chi(x,y;t),
	\label{equivariance of kernel:x} \\
	K^\chi(x,hy;t) & = & K^\chi(x,y;t) \rho^\chi(h^{-1}).
	\label{equivariance of kernel:y}
\end{eqnarray}
Now what we want to find is a path integral expression 
for the reduced kernel $ K^\chi $.

\section{Reduction of path integral}
To obtain a more concrete expression of the reduced kernel $ K^\chi(x,y;t) $,
we assume that $ M $ is a Riemannian manifold
on which the Lie group $ G $ acts with preserving the metric $ g_M $.
The Hamiltonian is 
\begin{equation}
	H = \dfrac12 \Delta_M + V(x),
	\label{D+V}
\end{equation}
where $ \Delta_M $ is the Laplacian of $ M $ and 
$ V(x) $ is an invariant potential function such that
$ V(gx) = V(x) $ for any $ x \in M $ and $ g \in G $.

The projection map $ p : M \to Q=M/G $ is equipped with 
stratification structure as we already discussed 
in the previous paper \cite{TI}.
Although the theory of stratification structure may look cumbersome
at a first glance,
it provides a natural language to describe quantization on
the inhomogeneous space $ M $.
Here we review the theory of stratification structure quickly.
The group $ G $ acts on $ M $ by isometry.
Let 
$ G_x := \{ g \in G \, | \, g x = x \} $ 
denote the isotropy group of $ x \in M $.
The orbit through $ x \in M $, 
$ {\cal O}_x = \{ gx \, | \, g \in G \} $, 
is diffeomorphic to a homogeneous space $ G/G_x $.
We call the tangent vector space
$ V_x = T_x {\cal O}_x $ a vertical subspace
and the orthogonal complement
$ H_x = (V_x)^\perp $ a horizontal subspace at $ x \in M $.
The vertical and horizontal projections,
$ P_V : T_x M \to V_x $ and
$ P_H : T_x M \to H_x $ respectively, are defined immediately.
A curve in $ M $ whose tangent vector is always in the horizontal subspace
is called a horizontal curve
and a map 
which transfers the beginning point of the horizontal curve to its end point
is called parallel transportation.
The dual spaces,
$ V_x^* = 
\{ \phi \in T_x^* M \, | \, \phi(v)=0, \, \forall v \in H_x \} $
and
$ H_x^* = 
\{ \phi \in T_x^* M \, | \, \phi(v)=0, \, \forall v \in V_x \} $,
are also defined.
Then the projection operators in the cotangent space,
$ P_V^* : T_x^* M \to V_x^* $ and
$ P_H^* : T_x^* M \to H_x^* $, are defined similarly.
Let $ \Lie $ denote the Lie algebra of the group $ G $.
Then the subgroup $ G_x $ is accompanied with the Lie subalgebra $ \Lie_x $.
The unitary representation $ \rho^\chi $ of the group $ G $
on the Hilbert space $ {\cal H}^\chi $
induces a representation $ \rho^\chi_* $ 
of the Lie algebra $ \Lie $ via differentiation.
The group action $ G \times M \to M $ also induces 
an infinitesimal transformation $ \Lie \times M \to TM $.
The tangent vectors generated by the Lie algebra 
define a linear map
$ \theta_x : \Lie \to T_x M $ at each point $ x \in M $.
Thus it is apparent that
$ \Kernel \theta_x = \Lie_x $ and $ \Image \theta_x = V_x $.
Then a quotient map 
$ \widetilde{\theta}_x : \Lie/\Lie_x \isoarrow V_x $ 
is an isomorphism.
The stratified connection form $ \omega $ is defined as
a linear map
\begin{equation}
	\omega_x = (\widetilde{\theta}_x)^{-1} \circ P_V 
	: 
	T_x M \to \Lie/\Lie_x.
	\label{connection form}
\end{equation}
We can also explain physical meanings of these geometric objects.
In the context of molecular mechanics \cite{TI}, 
the configuration space of the molecule is taken as the space $ M $
and the group of rotations $ SO(3) $ is taken as the group $ G $.
The vertical and horizontal spaces are called
the rotational and vibrational directions, respectively.
In this context the connection form represents
the angular velocity of the molecule
and the horizontal curve represents a vibrational motion
with zero angular velocity.

The path integral is usually introduced through the composition property
of the kernel
\begin{equation}
	K (x'',x; t+t')
	= 
	\int_M dx' \, K (x'',x';t') K (x',x;t).
	\label{composition}
\end{equation}
Repeating insertion of intermediate points, we obtain
\begin{equation}
	K (x',x; t)
	= 
	\int_M dx_1 \cdots dx_{n-1} \,
	K (x',x_{n-1}; \frac{t}{n}) \cdots
	K (x_2, x_1; \frac{t}{n})
	K (x_1, x  ; \frac{t}{n}).
	\label{repeated composition}
\end{equation}
{}For a short distance and a short time interval,
the kernel behaves asymptotically as
\begin{equation}
	K ( x_{i+1}, x_i; \Delta t ) 
	\sim
	\exp 
	\left[ i \Delta t
		\left(
		\frac{ \mbox{dist}^2 ( x_{i+1}, x_i ) }{ 2 \, (\Delta t)^2 }
		- V ( x_i )
		\right)
	\right]
	\label{asymptotic}
\end{equation}
and therefore 
a formal limit $ n \to \infty $ leads to the path integral expression
\begin{equation}
	K (x',x;t)
	=
	\int_x^{x'} [dx] 
	\exp
	\left[
		i \int_0^t ds 
		\left(
			\frac{1}{2}||\dot{x}(s)||^2 - V(x(s)) 
		\right)
	\right].
	\label{K_M}
\end{equation}
Although it is known \cite{DeWitte,Dowker}
that on a curved Riemannian manifold
the scalar curvature is added to the potential,
here we assume that it has already been included in the potential $ V $.

Let us $ \sigma : U \to M $ denote a local section over 
an open set $ U \subset Q $.
{}For each path $ x(s) $ $ (0 \le s \le t) $ in $ M $, 
a projected path $ q(s) := p(x(s)) $ in $ Q $ is defined.
If the path $ q(s) $ is contained in $ U $,
a path $ g(s) $ in $ G $ which satisfies
\begin{equation}
	x(s) = g(s) \sigma(q(s)),
	\quad
	0 \le s \le t,
\end{equation}
also exists.
We put 
$ x(0) = x  = g \sigma(q) $ and
$ x(t) = x' = g' \sigma(q') $.
The path integral (\ref{K_M}) is formally rewritten as
\begin{equation}
	K (x',x;t)
	=
	\int_q^{q'} [dq] \int_g^{g'} [dg] \,
	e^{i I [ g(\cdot) \sigma(q(\cdot)) ] },
	\label{path-int for K_M}
\end{equation}
where $ I $ denotes the action integral.
{}For a generic path $ x(s) $, 
the projected path $ q(s) = p(x(s)) $ is not contained 
in the single open set $ U \subset Q $.
Thus  we need to introduce a family of open covering sets 
$ \{ U_\alpha \} $
and a family of transformation functions $ \{ \phi_{\alpha \beta} \} $
to make the above equation (\ref{path-int for K_M}) meaningful
but they bring unnecessary cumbersomeness.
Hence we do not use them explicitly in this paper.
More careful treatment including the covering 
has been discussed in another paper \cite{TT1}.

When we insert an arbitrary smooth function $ \phi(s) $ taking values in $ G $
into the expression 
\begin{equation}
	x(s) = g(s) \phi(s)^{-1} \sigma(q(s)),
	\quad
	0 \le s \le t,
	\label{path}
\end{equation}
the path integral is left invariant;
\begin{equation}
	K (x',x;t)
	=
	\int_q^{q'} [dq] \int_{g \phi(0)}^{g' \phi(t)} [dg] \,
	e^{i I [ g(\cdot) \phi(\cdot)^{-1} \sigma(q(\cdot)) ] }.
	\label{inv path}
\end{equation}
Then we will choose the function $ \phi(s) $ 
to make a physical meaning of
the path integral expression of the reduced kernel $ K^\chi $
transparent.
We choose the function $ \phi(s) $ that makes
the curve $ \widetilde{x}(s) := \phi(s)^{-1} \sigma(q(s)) $
a horizontal curve starting from $ \widetilde{x}(0) = x $.
Namely, we demand that a tangent vector
\begin{equation}
	\frac{d}{ds} \widetilde{x}(s) 
	=
	\frac{d}{ds} 
	\biggl( \phi(s)^{-1} \sigma(q(s)) \biggr)
	=
	\phi^{-1}
	( 
	- \dot{\phi} \phi^{-1} \sigma
	+ \dot{\sigma}
	)
\end{equation}
has a vanishing vertical component. 
This requirement is equivalent to a differential equation
\begin{equation}
	- \dot{\phi} {\phi}^{-1} \sigma
	+ P_V (\dot{\sigma})
	=
	( 
		- \dot{\phi} {\phi}^{-1}
		+ \omega(\dot{\sigma}) 
	) \sigma
	= 0
	\label{ODE1}
\end{equation}
for $ \phi(s) $ with an initial condition $ \phi(0)^{-1} = g $.
Here we have used the stratified connection form $ \omega $,
which was defined in (\ref{connection form}).
Equation (\ref{ODE1}) is rewritten as
\begin{equation}
	\dot{\phi}(s)
	= \omega( \dot{\sigma}(s) ) \, \phi(s)
	\label{ODE2}
\end{equation}
and its solution is formally written as
\begin{equation}
	\phi(t)
	=
	W_\sigma (t) \phi(0)
	=
	W_\sigma (t) g^{-1},
	\qquad
	W_\sigma (t)
	= 
	{\cal P} e^{\int_0^t ds \, \omega(\dot{\sigma}) },
	\label{solution}
\end{equation}
where $ {\cal P} $ is a symbol indicating the path-ordered product.

In the tangent vector of the curve (\ref{path})
\begin{equation}
	\frac{d}{ds} x
	= 
	\frac{dg}{ds} \, \widetilde{x}
	+ g \frac{d \widetilde{x}}{ds},
	\label{tangent}
\end{equation}
the first term is a vertical vector 
and the second term is a horizontal one
according to the definition of $ \phi(s) $.
Then the norm of the tangent vector (\ref{tangent}) with respect to the metric
$ g_M $ is given as
\begin{equation}
	|| \dot{x} ||^2
	= 
	|| P_V (\dot{x}) ||^2 + || P_H (\dot{x}) ||^2
	= 
	|| \dot{g} ||^2 + || \dot{q} ||^2.
	\label{decompo of dot x}
\end{equation}
The last line should be understood as the definitions of
$ || \dot{g} ||^2 $ and $ || \dot{q} ||^2 $.
Putting Eqs.
(\ref{equivariant kernel}),
(\ref{K_M}),
(\ref{path-int for K_M}) and
(\ref{inv path})
together we obtain
\begin{eqnarray}
	K^{\chi} ( x', x; t )
& = &
	\int_G dh \, \rho^\chi(h) \, 
	K ( h^{-1} x', x; t )
	\nonumber \\
& = &
	\int_G dh \, \rho^\chi(h) 
	\int_{x}^{h^{-1} x'} [dx] 
	\, e^{ i I[ x(\cdot) ] }
	\nonumber \\
& = &
	\int_G dh \, \rho^\chi(h) 
	\int_{q}^{q'} [dq]  
	\int_g^{h^{-1} g'} [dg]
	\, e^{ i I[ g(\cdot) \sigma(q(\cdot)) ] }
	\nonumber \\
& = &
	\int_{q}^{q'} [dq]
	\int_G dh 
	\, \rho^\chi(h) 
	\int_{g \phi(0)}^{ h^{-1} g' \phi(t) } [dg] 
	\, e^{ i I[ g(\cdot) \phi(\cdot)^{-1} \sigma(q(\cdot)) ] }.
\end{eqnarray}
By changing the integral variable $ h $ to $ g' \phi(t) h $
and using the decomposition (\ref{decompo of dot x}) of
the kinetic energy $ || \dot{x} ||^2 $,
and furthermore by substituting the solution (\ref{solution})
for $ \phi(t) $ with $ \phi(0) = g^{-1} $, we obtain
\begin{eqnarray}
&&
	K^{\chi} ( x', x; t )
	\nonumber \\
& = &
	\int_{q}^{q'} [dq]
	\int_G dh 
	\, \rho^\chi( g' \phi(t) h ) 
	\int_e^{h^{-1}} [dg] 
	\, e^{
		i \int ds 
		\{ \frac12 || \dot{g} ||^2 + \frac12 || \dot{q} ||^2 - V(q) \}
		}
	\nonumber \\
& = &
	\int_{q}^{q'} [dq]
	\, \rho^\chi( g' W_\sigma (t) g^{-1} )
	\left\{
		\int_G dh 
		\, \rho^\chi( h )
		\int_e^{h^{-1}} [dg] 
		\, e^{ \frac{i}{2} \int ds || \dot{g} ||^2 }
	\right\}
	\, e^{
		i \int ds 
		\{ \frac12 || \dot{q} ||^2 - V(q) \}
	}.
	\qquad \quad
	\label{reduced path integral}
\end{eqnarray}
We call the element
$ \rho^\chi( g' W_\sigma (t) g^{-1} ) \in \End ( {\cal H}^\chi ) $
a holonomy factor for the following reason.

A geometrical meaning of the holonomy factor
$ \rho^\chi( g' W_\sigma (t) g^{-1} ) $ is easy to understand.
Two curves $ x(s) $ and $ \widetilde{x}(s) $ have the end points
\begin{eqnarray}
	&&
	x(t) 
	= 
	x' 
	= g' \sigma(q(t)),
	\label{x'} \\
	&&
	\widetilde{x}(t)
	= 
	\phi(t)^{-1} \sigma(q(t)) 
	= g W_\sigma(t)^{-1} \sigma(q(t)).
	\label{tilde x'}
\end{eqnarray}
Then they are related as
\begin{equation}
	x(t) 
	 = ( g' W_\sigma (t) g^{-1} ) \cdot \widetilde{x}(t).
	\label{difference}
\end{equation}
Hence the group element $ g' W_\sigma (t) g^{-1} $ indicates
how much the end point $ x(t) $ differs from
the parallelly transported point $ \widetilde{x}(t) $.
Even if the projected curve $ q(s) $ forms a closed loop
as $ q(t) = q(0) $, 
\begin{equation}
	\widetilde{x}(t) 
	= g W_\sigma(t)^{-1} \sigma(q(t))
	= g W_\sigma(t)^{-1} \sigma(q(0))
	= ( g W_\sigma(t)^{-1} g^{-1} ) \cdot \widetilde{x}(0),
	\label{holonomy}
\end{equation}
thus the horizontal curve does not close in general.
The group element $ g W_\sigma(t)^{-1} g^{-1} $ is called a holonomy
associated with the loop $ q(s) $ in proper words of
differential geometry of connection.
However we also loosely call the element 
$ \rho^\chi( g' W_\sigma (t) g^{-1} ) $ the holonomy factor.

A physical meaning of the holonomy factor is interesting;
if the group $ G $ is the Abelian group $ U(1) $,
we can write the connection form as $ \omega = i A $
with a real valued one-form $ A $.
The holonomy factor in (\ref{solution}) is then written as
\begin{equation}
	W_\sigma (t)
	=
	\exp \left[ i \! \int A_j dx^j \right],
\end{equation}
thus it brings the gauge potential $ A $ to the Lagrangian effectively.
{}For a general non-Abelian group $ G $ the holonomy factor
is to be understood as a coupling of the system
with the induced gauge field $ \omega $.

The factor in the brace of (\ref{reduced path integral}),
\begin{equation}
	\int_G dh 
	\, \rho^\chi( h )
	\int_e^{h^{-1}} [dg] 
	\, e^{ \frac{i}{2} \int ds || \dot{g} ||^2 },
	\label{G}
\end{equation}
is still left less understood.
Then we examine it carefully in the subsequent section.

\section{Rotational energy amplitude}
To explain the above path integral expression (\ref{G})
we have to prepare more notations.
The Riemannian manifold $ M $ is equipped with
a volume form $ v_M $ associated with the metric $ g_M $.
The metric $ g_M = g_V + g_H $ is decomposed
according to the decomposition of the tangent space
$ T_x M = V_x \oplus H_x $.
Then the Laplacian $ \Delta_M $ is also decomposed in a similar manner;
suppose that 
$ \psi(x) \in C^\infty_c (M; {\cal H}^\chi) $ 
is a $ C^\infty $ function
taking values in $ {\cal H}^\chi $ with a compact support.
Then with the use of projectors $ P_V^* $ and $ P_H^* $ on $ T_x^* M $,
we define vertical and horizontal components of the Laplacian through
\begin{eqnarray}
	\int_M || d \psi ||^2 \, v_M 
	& = &
	\int_M \bra \psi, \, \Delta_M \psi \ket \, v_M,
	\label{delta_M} \\
	\int_M || P_V^* (d \psi) ||^2 \, v_M 
	& = &
	\int_M \bra \psi, \, \Delta_V \psi \ket \, v_M,
	\label{delta_V} \\
	\int_M || P_H^* (d \psi) ||^2 \, v_M 
	& = &
	\int_M \bra \psi, \, \Delta_H \psi \ket \, v_M.
	\label{delta_H}
\end{eqnarray}
Notice that in our convention the Laplacians
$ \Delta_M, \, \Delta_V, \, \Delta_H, $ are nonnegative operators.

Now we can write the path integral in (\ref{G}) as
\begin{equation}
	K_V ( g \widetilde{x}(t), \widetilde{x}(0); t )
	=
	\bra g \widetilde{x}(t) | 
	e^{ - \frac{i}{2} t \Delta_V } 
	| \widetilde{x}(0) \ket
	=
	\int_e^{g} [dg] 
	\, e^{ \frac{i}{2} \int ds || \dot{g} ||^2 },
	\label{meaning of the kernel}
\end{equation}
which has a subtle meaning.
At each point $ q \in Q $, a fiber $ F = p^{-1}(q) $ exists.
If any point $ x \in p^{-1}(q) $ is chosen as a reference point,
a map
\begin{equation}
	\varepsilon(x) : G/G_x \to {\cal O}_x = F;
	\qquad
	[g] \mapsto gx
	\label{epsilon_x}
\end{equation}
is defined and turns out to be a diffeomorphism.
Thus we have a family of parametrized fibers
$ \{ F_s = p^{-1} (q(s)) \}_{0 \le s \le t } $
over the projected curve $ q(s) $.
Now a diffeomorphism between fibers $ F_s $ and $ F_0 $
is defined by a map
$ \varepsilon(\widetilde{x}(0)) \circ \varepsilon(\widetilde{x}(s))^{-1} :
F_s \isoarrow F_0 $
using the points $ \widetilde{x}(0) $ and $ \widetilde{x}(s) $ as references.
Hence the family of fibers
$ \{ F_s \}_{0 \le s \le t } $
can be identified with the direct product space 
$ F_0 \times [0,t] $.
Moreover, each fiber $ F_s $ is equipped with a Riemannian metric $ g_s $
which is defined by restricting the original metric $ g_M $ to the fiber.
The metric $ g_s $ coincides also with the vertical metric $ g_V $
restricted to $ F_s $.
It is to be noted that
the different fibers $ (F_s, g_s ) $ and $ (F_{s'}, g_{s'} ) $ 
are not necessarily isometric for $ s \ne s' $.
The metric $ g_s $ defines also the Laplacian $ \Delta_s $ of each fiber $ F_s $.
Then the kernel 
$ K_V ( \cdot, \cdot, s ) : F_s \times F_0 \to \C $
in (\ref{meaning of the kernel}) is the fundamental solution of 
the Schr{\"o}dinger equation
\begin{equation}
	\frac{\partial}{\partial s} K_V (x, x_0; s)
	= - \frac{i}{2} \Delta_s^x K_V (x, x_0; s)
	\label{K_V}
\end{equation}
with the initial condition $ K_V (x, x_0; 0) = \delta_F (x, x_0) $.
This argument explains a meaning of 
the first two terms of Eq.(\ref{meaning of the kernel}).
Since in the context of molecular mechanics
the fiberwise directions represent rotational motions,
it seems suitable to call the kernel $ K_V (x, x_0; s) $
the rotational energy amplitude.

Next we turn to the third term of (\ref{meaning of the kernel}),
\begin{equation}
	\int_e^g [dg] 
	\, e^{ \frac{i}{2} \int ds || \dot{g} ||^2 }.
	\label{path over G}
\end{equation}
It should be understood as an integration over paths 
$ g(s) \widetilde{x}(s) $,
which vary by multiplying various $ g(s) $ 
on the fixed reference path $ \widetilde{x}(s) $.
By repeating the usual argument which leads to the path integral (\ref{K_M}),
we arrive at the expression (\ref{path over G}) 
for the kernel $ K_V (x, x_0; s) $.

Moreover, the rotational energy amplitude in (\ref{G}), 
\begin{eqnarray}
	K_V^\chi ( \widetilde{x}(t), \widetilde{x}(0); t )
	& := &
	\int_G dh 
	\, \rho^\chi( h )
	K_V ( h^{-1} \widetilde{x}(t), \widetilde{x}(0); t )
	\nonumber \\
	& = &
	\int_G dh 
	\, \rho^\chi( h )
	\int_e^{h^{-1}} [dg] 
	\, e^{ \frac{i}{2} \int ds || \dot{g} ||^2 }
	\label{K_V^chi}
\end{eqnarray}
is to be understood as
the kernel restricted to equivariant functions as stated in
(\ref{equivariant kernel}).
Accordingly we will show that the equivariant kernel (\ref{K_V^chi}) 
can be reduced to a much simpler expression below.

We now examine the action of the vertical Laplacian $ \Delta_V $ on 
an equivariant function $ \psi(x) $.
When $ \psi(x) $ is equivariant, for any $ X \in \Lie $ we have
\begin{equation}
	\theta(X) \psi = \rho_*^\chi(X) \psi,
\end{equation}
which is derived from (\ref{equivariant}) by differentiation.
If we put $ X = \omega(v) $ for some $ v \in T_x M $,
we have $ \theta( \omega(v) ) = P_V (v) $ according to (\ref{connection form}).
Hence we obtain
\begin{equation}
	P_V^* (d \psi) = \rho_*^\chi( \omega ) \psi.
\end{equation}
Therefore, Eq.(\ref{delta_V}) is rewritten as
\begin{eqnarray}
	\int_M || P_V^* (d \psi) ||^2 \, v_M 
& = &
	\int_M || \rho_*^\chi( \omega ) \psi ||^2 \, v_M 
	\nonumber \\
& = &
	\int_M g_M^{-1} 
	\bra 
		\rho_*^\chi( \omega ) \psi,
		\rho_*^\chi( \omega ) \psi
	\ket \, v_M 
	\nonumber \\
& = &
	- \int_M g_M^{-1} 
	\bra 
		\psi,
		\rho_*^\chi( \omega ) \rho_*^\chi( \omega ) \psi
	\ket \, v_M,
	\label{D_V}
\end{eqnarray}
where $ g_M^{-1} $ is a metric on the cotangent bundle.
Here notice that $ \rho_*^\chi $ takes values in anti-hermitian elements
of $ \End ( {\cal H}^\chi ) $.
Combining the metric $ g_M^{-1}(x) : T^*_x M \otimes T^*_x M \to \R $
with the connection form $ \omega_x \in ( \Lie / \Lie_x ) \otimes T^*_x M $,
let us define an element of the tensor product of the Lie algebra
\begin{equation}
	{\mit{\Lambda}}_x :=
	- g_M^{\, -1}(x) \circ ( \omega_x \otimes \omega_x )
	\in ( \Lie / \Lie_x ) \otimes ( \Lie / \Lie_x ),
	\label{rotational energy operator}
\end{equation}
which we call the rotational energy operator.
The representation of the Lie algebra
$ \rho_*^\chi : \Lie \to \End ( {\cal H}^\chi ) $
can be extended to a representation of the universal envelop algebra
$ \rho_*^\chi : {\cal U}(\Lie) \to \End ( {\cal H}^\chi ) $.
Thus the value of 
$ {\mit{\Lambda}}_x \in \Lie \otimes \Lie $
is mapped to
$ \rho_*^\chi ( {\mit{\Lambda}}_x ) \in \End ( {\cal H}^\chi ) $.
It is also apparent that 
$ \rho_*^\chi ( {\mit{\Lambda}}_x ) $
is a nonnegative operator from the definition.
Then from (\ref{delta_V}), (\ref{D_V}) and (\ref{rotational energy operator}),
we conclude that the vertical Laplacian $ \Delta_V $
is reduced to an algebraic operator
when it acts on equivariant functions as
\begin{equation}
	\Delta_V \psi(x) = \rho_*^\chi ( {\mit{\Lambda}}_x ) \psi(x).
	\label{reduction of D_V}
\end{equation}
Then we can write the equivariant vertical kernel (\ref{K_V^chi}) as
\begin{equation}
	\int_G dh 
	\, \rho^\chi( h )
	\bra h^{-1} \widetilde{x}(t) | 
	e^{ - \frac{i}{2} t \Delta_V } 
	| \widetilde{x}(0) \ket
	=
	{\cal P} \exp
	\left[
		- \frac{i}{2} \int_0^t ds \,
		( \rho_*^\chi \circ {\mit{\Lambda}} ) ( \widetilde{x}(s) )
	\right].
	\label{fiber kernel}
\end{equation}
If we put
\begin{equation}
	R(t)
	:=
	{\cal P} \exp
	\left[
		- \frac{i}{2} \int_0^t ds \,
		{\mit{\Lambda}} ( \widetilde{x}(s) )
	\right],
	\label{R}
\end{equation}
which takes values in the universal envelop algebra $ {\cal U}(\Lie) $,
then the equivariant kernel (\ref{fiber kernel}) is represented as
$ \rho^\chi_* (R(t)) $.
{}Finally, by substituting this into (\ref{reduced path integral}),
we arrive at the expression of the reduced kernel
\begin{eqnarray}
	K^\chi(x',x;t)
& = &
	\int_{q}^{q'} [dq]
	\, \rho^\chi( g' W_\sigma (t) g^{-1} )
	\, \rho_*^\chi( R(t) )
	\, e^{
		i \int ds 
		\{ \frac12 || \dot{q} ||^2 - V(q) \}
	}
	\nonumber \\
& = &
	\int_{q}^{q'} [dq] \, 
	\rho^\chi
		\left( 
			g' \,
			{\cal P} \exp 
			\left[ \int_0^t ds \, \omega(\dot{\sigma}) \right]
			g^{-1} 
		\right)
	\nonumber \\
&&	\qquad \quad
	\rho_*^\chi 
		\left(
			{\cal P} \exp
			\left[
				- \frac{i}{2} \int_0^t ds \,
				{\mit{\Lambda}} ( \widetilde{x}(s) )
			\right]
		\right)
	\nonumber \\
&&	\qquad \qquad
	\exp 
	\left[ 
		i \int_0^t ds 
		\left\{ 
			\frac12 || \dot{q} ||^2 - V(q) 
		\right\} 
	\right],
	\label{main result}
\end{eqnarray}
which is a main result of this paper.
The remaining phase factor
$ \exp [ i \int ds \{ \frac12 || \dot{q} ||^2 - V(q) \} ] $
is called the vibrational energy amplitude 
in the context of molecular mechanics.

\section{Singular points and boundary condition}
Here let us mention characteristic behavior of the equivariant kernel
(\ref{equivariant kernel}) at a singularity.
Suppose that a point $ x \in M $ admits a nontrivial isotropy group
$ G_x \ne \{ e \} $.
Then the equivariance of the kernel (\ref{equivariance of kernel:x}) 
leads to the invariance
\begin{equation}
	K^\chi(x,y;t)
	= 
	K^\chi(hx,y;t)
	= 
	\rho^\chi(h) K^\chi (x,y;t),
	\qquad h \in G_x.
\end{equation}
Then the value of $ K^\chi(x,y;t) $ is restricted to the subspace of
invariant vectors
\begin{equation}
	({\cal H}^\chi)^{G_x} := 
	\{ v \in {\cal H}^\chi \, | \, \rho^\chi(h) v=v, \, 
	\forall h \in G_x \}.
\end{equation}
Let $ dh $ denote an normalized invariant measure of $ G_x $.
Then let us introduce an operator
\begin{equation}
	B^\chi (x) := 
	\int_{G_x} dh \, \rho^\chi (h),
	\label{B}
\end{equation}
which is an element of $ \End( {\cal H}^\chi ) $.
It is easily verified that $ B^\chi(x) $ is a projection operator onto
$ ({\cal H}^\chi)^{G_x} $.
Thus the value of the kernel $ K^\chi(x,y;t) $ automatically satisfies
\begin{equation}
	B^\chi(x) K^\chi(x,y;t)
	= 
	K^\chi(x,y;t)
\end{equation}
and in a composition relation similar to (\ref{composition})
\begin{equation}
	K^\chi(x'',x; t+t')
	= 
	\int_M dx' \, K^\chi(x'',x';t') K^\chi(x',x;t),
	\label{composition of K^chi}
\end{equation}
we can freely insert the projection operator as
\begin{equation}
	K^\chi(x'',x; t+t')
	= 
	\int_M dx' \, K^\chi(x'',x';t') B^\chi(x') K^\chi(x',x;t).
\end{equation}
When the path runs through a singular point $ x' $, at which
the dimension of $ G_{x'} $ abruptly increases,
the rank of the projection operator $ B^\chi (x') $ decreases.
This gives a kind of boundary condition to the kernel
when the path hits a singular point.

\section{Example}
Let us explore a simple application of the above developed formalism.
As an example
we take the Euclidean space $ \R^2 $ with the standard metric 
$ g_M = dx^2 + dy^2 = dr^2 + r^2 d \theta^2 $ as $ M $.
The group $ G = SO(2) $ acts on $ \R^2 $ and defines a quotient space
$ Q = \R^2 / SO(2) = \Rplus $.
The group action
\begin{equation}
	SO(2) \times \R^2 \to \R^2;
	\quad
	\left(
		\begin{array}{rr}
		\cos \phi & - \sin \phi \\
		\sin \phi &   \cos \phi 
		\end{array}
	\right)
	\left(
		\begin{array}{c}
		x \\ y
		\end{array}
	\right)
	\label{action of SO(2)}
\end{equation}
induces the action of the Lie algebra
\begin{equation}
	\LieSO(2) \times \R^2 \to T\R^2;
	\quad
	\left(
		\begin{array}{rr}
		0    & - \phi \\
		\phi &   0      
		\end{array}
	\right)
	\left(
		\begin{array}{c}
		x \\ y
		\end{array}
	\right),
	\label{action of LieSO(2)}
\end{equation}
which also defines a map
\begin{equation}
	\theta : \LieSO(2) \times \R^2 \to T\R^2;
	\quad
	\left(
	\left(
		\begin{array}{rr}
		0    & - \phi \\
		\phi &   0      
		\end{array}
	\right),
	\,
	\left(
		\begin{array}{c}
		r \cos \theta \\ 
		r \sin \theta
		\end{array}
	\right)
	\right)
	\mapsto
	\phi \frac{\partial}{\partial \theta}.
	\label{theta of LieSO(2)}
\end{equation}%
Then the stratified connection (\ref{connection form})
now becomes a one-form taking values in $ \LieSO(2) $,
\begin{equation}
	\omega =
	\left(
		\begin{array}{rr}
		0 & -1 \\
		1 &  0   
		\end{array}
	\right)
	d \theta.
	\label{omega}
\end{equation}
The vertical and horizontal components of the metric are given as
$ g_V = r^2 d \theta^2 $ and $ g_H = dr^2 $, respectively.
In the dual space the metric is given as
\begin{equation}
	(g_M)^{-1} = 
	\dfrac{\partial}{\partial r} \otimes
	\dfrac{\partial}{\partial r}
	+
	\dfrac{1}{r^2} 
	\dfrac{\partial}{\partial \theta} \otimes
	\dfrac{\partial}{\partial \theta}.
\end{equation}
The rotational energy operator (\ref{rotational energy operator})
is now calculated as
\begin{equation}
	{\mit{\Lambda}}
	= -(g_M)^{-1} \circ ( \omega \otimes \omega ) 
	=
	- \dfrac{1}{r^2}
	\left(
		\begin{array}{rr}
		0 & -1 \\
		1 &  0   
		\end{array}
	\right)
	\otimes
	\left(
		\begin{array}{rr}
		0 & -1 \\
		1 &  0   
		\end{array}
	\right).
\end{equation}
Any irreducible unitary representation of $ SO(2) $ is one-dimensional.
It is labeled by an integer $ n \in \Z $ and defined by
\begin{equation}
	\rho_n : SO(2) \to U(1);
	\quad
	\left(
		\begin{array}{rr}
		\cos \phi & - \sin \phi \\
		\sin \phi &   \cos \phi 
		\end{array}
	\right)
	\mapsto
	e^{i n \phi}.
	\label{rep of SO(2)}
\end{equation}
Thus the differential representation of the Lie algebra of $ SO(2) $
is defined by
\begin{equation}
	(\rho_n)_* : \LieSO (2) \to \LieU (1);
	\quad
	\left(
		\begin{array}{rr}
		0 & - \phi \\
		\phi &  0 
		\end{array}
	\right)
	\mapsto
	i n \phi.
	\label{rep of LieSO(2)}
\end{equation}
The rotational energy operator is then represented as
\begin{equation}
	(\rho_n)_* ( {\mit{\Lambda}} )
	=
	- \dfrac{1}{r^2} ( in )^2
	=
	\dfrac{n^2}{r^2}.
\end{equation}
The origin $ r=0 $ admits a nontrivial isotropy group $ G_0 = SO(2) $.
The boundary projection operator (\ref{B}) becomes
\begin{equation}
	B_n (0) = 
	\int_0^{2 \pi} \frac{d \phi}{2 \pi} \, e^{in \phi} = 
	\delta_{n0}.
\end{equation}
If we take the map $ r \mapsto (r,0) $ 
as a section $ \sigma : \Rplus \to \R^2 $
then the pullback of the connection identically vanishes,
$ \sigma^* \omega = 0 $,
and the reduced path integral (\ref{main result}) is now given by
\begin{equation}
	K_n(r',r;t)
	= 
	\int_{r}^{r'} [r \,dr] 
	B_n [r]
	\exp
	\left[
		i \int_0^t ds
		\left\{
		\frac12 \dot{r}^2 - \frac{n^2}{2 r^2} - V(r)
		\right\}
	\right],
	\label{path integral on R+}
\end{equation}
where the factor $ B_n[r] $ represents the boundary condition at $ r=0 $.
If $ n \ne 0 $ and $ r(s)=0 $ at some time $ s $ between $ 0 \le s \le t $,
then $ B_n[r] = 0 $.
Otherwise, $ B_n[r] = 1 $.

We can compare this result (\ref{path integral on R+}) with
a reduced kernel obtained from an exact kernel
for the specific case of a free particle with the Hamiltonian
\begin{equation}
	H 
	= \dfrac12 \Delta 
	= - \dfrac12 
	\left( 
		\dfrac{\partial^2}{\partial x^2} +
		\dfrac{\partial^2}{\partial y^2}  
	\right).
\end{equation}
The exact kernel on $ \R^2 $ is
\begin{equation}
	K((x_2,y_2), (x_1,y_1); t)
	=
	\frac{1}{2 \pi i t} 
	\exp
	\left[
		\frac{i}{2t}
		\left\{
			(x_2 - x_1)^2 + (y_2 - y_1)^2 
		\right\}
	\right].
\end{equation}
If we put 
$ z_j = x_j + i y_j = r_j \, e^{i \theta_j} $ $ (j=1,2) $,
it is rewritten as
\begin{eqnarray}
	K(z_2, z_1; t)
	& = &
	\frac{1}{2 \pi i t} 
	\exp
	\left[
		\frac{i}{2t}
		| z_2 - z_1 |^2
	\right]
	\nonumber \\
	& = &
	\frac{1}{2 \pi i t} 
	\exp
	\left[
		\frac{i}{2t}
		\left\{
		r_2^2 + r_1^2 - 2 r_1 r_2 \cos ( \theta_2 - \theta_1)
		\right\}
	\right].
\end{eqnarray}
Then the reduced kernel is explicitly calculated as
\begin{eqnarray}
	K_n (z_2, z_1; t)
& = &
	\int_0^{2 \pi} \frac{d \phi}{2 \pi} \,
	\rho_n( e^{i \phi} )
	K( e^{-i \phi} \, z_2, z_1; t)
	\nonumber \\
& = &
	\frac{1}{2 \pi i t} 
	\int_0^{2 \pi} \frac{d \phi}{2 \pi} \,
	e^{i n \phi}
	\exp
	\left[
		\frac{i}{2t}
		\left\{
		r_2^2 + r_1^2 - 2 r_1 r_2 
		\cos ( \theta_2 - \theta_1 - \phi)
		\right\}
	\right]
	\nonumber \\
& = &
	\frac{1}{2 \pi i t} 
	e^{ i ( r_2^2 + r_1^2 )/(2t) }
	e^{in (\theta_2 - \theta_1) }
	\int_0^{2 \pi} \frac{d \phi}{2 \pi} \,
	e^{i n \phi}
	\exp
	\left[
		- \frac{i}{t} r_1 r_2 \cos \phi
	\right]
	\nonumber \\
& = &
	\frac{1}{2 \pi i t} 
	e^{ i ( r_2^2 + r_1^2 )/(2t) }
	e^{in (\theta_2 - \theta_1 - \pi/2) }
	J_n 
	\left( \frac{1}{t} r_1 r_2 \right),
\end{eqnarray}
where $ J_n(x) $ denotes the $n$-th Bessel function.
Asymptotic behavior of the Bessel function for a short time interval 
$ t $ such that $ t / r_1 r_2 << 1 $ is given by
\begin{equation}
	J_n 
	\left( \frac{r_1 r_2}{t} \right)
	\sim
	\sqrt{ \frac{2 t}{\pi r_1 r_2} }
	\exp
	\left[
		- i 
		\left(
		\frac{r_1 r_2}{t} + \frac{n^2 t}{2 r_1 r_2} - \frac{n \pi}{2}
		\right)
	\right].
	\label{approx}
\end{equation}
The derivation of this formula is shown in the appendix. 
Thus for the short interval the reduced kernel behaves as
\begin{equation}
	K_n (z_2, z_1; t)
	\sim 
	\frac{1}{ \pi i \sqrt{ 2 \pi t r_1 r_2} }
	e^{in (\theta_2 - \theta_1)}
	\exp
	\left[
		i t
		\left\{
		\frac{(r_2 - r_1)^2}{2 \, t^2} - \frac{n^2}{2 \, r_1 r_2}
		\right\}
	\right],
\end{equation}
which reproduces the integrand of the path integral (\ref{path integral on R+}).

\section{Conclusion}
Now let us summarize our discussion.
We began with a review of the method of reduction of quantum systems.
We assumed that
the original system was defined in terms of the Riemannian manifold $ M $
and that the isometric transformations of $ M $ formed 
the compact Lie group $ G $.
Then the projection map $ p : M \to Q=M/G $ was equipped with
stratification structure, which is a generalization of
principal fiber bundle structure.
Our main purpose was to give a path integral expression 
on the quotient space $ Q $
to the time-evolution operator of the reduced system.
The reduced path integral (\ref{main result})
was factorized into three parts;
the rotational energy amplitude,
the vibrational energy amplitude,
and the holonomy factor.
The rotational energy amplitude represented 
the integration over the fiber directions,
while the vibrational energy amplitude
represented the integration over the directions perpendicular to the fiber.
These perpendicular directions defined the connection,
and when they were non-integrable, the holonomy factor arose.

At the singular point which was characterized by larger symmetry
than the neighboring points, the amplitude was restricted 
to values invariant under the symmetry operations,
and therefore the boundary value projection operator 
was automatically introduced.
As a simple example, we applied our formalism to the quantum system
on $ \R^2 $ with symmetry specified by $ SO(2) $.
{}For the case of a free particle,
our result was compared with the exact result,
and then we confirmed their agreement.

Finally, we would like to mention remaining problems.
{}For further application,
the quantum system on the sphere $ SU(2)/U(1) $
and the one on the meridian line $ U(1) \backslash SU(2) / U(1) $
are also interesting to study the role of the boundary conditions.
The quantum system on the adjoint orbit space $ G / \mbox{Ad} G $ 
of a Lie group $ G $ is also interesting for physical application
since it has a role as a toy model of 
the gauge field theory as other authors \cite{Rajeev,Mickelsson} discussed.
Hence it is expected that our method of reduction of quantum systems
is useful for gauge invariant analysis of the field theory, 
which is a subject attracting attention recently 
in high energy physics \cite{LM1,LM2}.

Although the path integral formalism presented here 
is a method to treat quantum systems, 
it is naturally applicable to stochastic processes, too.
Statistical problems concerning the shape space $ Q = M/G $
have been studied by Kendall \cite{statistics of shape space}
and other people.
The Schr{\"o}dinger equation is replaced 
by the diffusion equation in the context of stochastic problems,
and the Wiener integral could be also reduced 
in a similar way if it admits symmetry.

{}For a strong attractive potential, 
for example, the inverse-square potential $ V=-k r^{-2} $, 
the particle falls into the origin within a finite time 
in the context of classical mechanics.
It is an interesting question \cite{KT,Landsman}
to ask what kind of phenomenon in quantum mechanics 
is corresponding to the finite-falling particle.
Facing such a problem, 
it is expected that we need to modify the boundary reflection condition
but we do not have a definite answer yet.

\section*{Acknowledgments}
The author wishes to thank Professor Iwai for 
valuable comments and criticisms.
He gratefully acknowledges also encouraging discussions
with Tsutsui.
This work was supported by a Grant-in-Aid for Scientific Research from
the Ministry of Education, Science and Culture of Japan.

\renewcommand{\thesection}{Appendix A.}
\renewcommand{\theequation}{A.\arabic{equation}}
\section{Asymptotic formula of Hankel}
The asymptotic behavior of the Bessel function $ J_n(x) $ is given by
the following equations: if two functions
\begin{eqnarray}
	A_N (x) & = &
		1 +
		\sum_{r=1}^{[N/2]} (-1)^r
		\frac{(4n^2-1^2) (4n^2-3^2) \cdots [4n^2-(4r-1)^2]}%
		{ (2r)! \, (8x)^{2r} },
	\label{A_N} \\
	B_N (x) & = &
		\sum_{r=0}^{[(N-1)/2]} (-1)^r
		\frac{(4n^2-1^2) (4n^2-3^2) \cdots [4n^2-(4r+1)^2]}%
		{ (2r+1)! \, (8x)^{2r+1} },
	\label{B_N}
\end{eqnarray}
are introduced, 
for $ |x| >> n, 1 $ the Bessel function is expanded as
\begin{equation}
	J_n(x) \sim
	\sqrt{ \frac{2}{\pi x} }
	\left\{
		A_N(x) \cos \left( x - \frac{(2n+1) \pi}{4} \right) -
		B_N(x) \sin \left( x - \frac{(2n+1) \pi}{4} \right)
	\right\},
	\label{Hankel}
\end{equation}
which is called Hankel's asymptotic expansion formula of the Bessel function.
Now we put
\begin{equation}
	\omega := x - \frac{(2n+1) \pi}{4},
	\qquad
	\varepsilon := B_0 (x) =
		\frac{ 4n^2-1^2 }{ 8x } \sim
		\frac{  n^2 }{ 2x },
	\label{epsilon}
\end{equation}
then the Bessel function is approximated as
\begin{eqnarray}
	J_n(x) 
& \sim &
	\sqrt{ \frac{2}{\pi x} }
	\left\{
		A_0(x) \cos \omega - B_0(x) \sin \omega
	\right\}
	\nonumber \\
& = &
	\sqrt{ \frac{2}{\pi x} }
	\left\{
		\cos \omega - \varepsilon \sin \omega
	\right\}
	\nonumber \\
& \sim &
	\sqrt{ \frac{2}{\pi x} }
		\cos ( \omega + \varepsilon ).
	\label{approximate Bessel}
\end{eqnarray}
Thus the Bessel function that we encountered in this paper
is approximated as
\begin{equation}
	J_n 
	\left( \frac{r_1 r_2}{t} \right)
	\sim
	\sqrt{ \frac{2 t}{\pi r_1 r_2} }
	\Re
	\exp
	\left[
		- i 
		\left(
		\frac{r_1 r_2}{t} + \frac{n^2 t}{2 r_1 r_2} - \frac{(2n+1) \pi}{4}
		\right)
	\right],
\end{equation}
which justifies Eq. (\ref{approx}) in the body of this paper.

\baselineskip 5mm 


\begin{thebibliography}{99}
\bibitem{Arnold}
	V. I. Arnold,
	{\sl Mathematical methods of classical mechanics},
	translated by K. Vogtmann and A. Weinstein
	(Springer-Verlag, 1989),
	2nd ed. 
\bibitem{Marsden-Weinstein}
	J. Marsden and A. Weinstein,
	{\sl Rep. Math. Phys.} {\bf 5}, 121 (1974).
\bibitem{Guichardet}
	A. Guichardet,
	{\sl Ann. Inst. Henri Poincar{\'e}} {\bf 40}, 329 (1984).
\bibitem{Iwai1}
	T. Iwai,
	{\sl Ann. Inst. Henri Poincar{\'e}} {\bf 47}, 199 (1987).
\bibitem{Iwai3}
	T. Iwai,
	{\sl J. Math. Phys.} {\bf 28}, 1315 (1987).
\bibitem{Montgomery}
	R. Montgomery,
	{\sl Commun. Math. Phys.} {\bf 128}, 565 (1990).
\bibitem{Davis paper}
	M. Davis,
	{\sl Pacific J. Math.} {\bf 77}, 315 (1978).
\bibitem{TI}
	S. Tanimura and T. Iwai,
	{\sl J. Math. Phys.} {\bf 41}, 1814 (2000).
\bibitem{Wren}
	K. K. Wren,
	{\sl J. Geom. Phys.} {\bf 24}, 173 (1998).
\bibitem{Schulman}
	L. S. Schulman, 
	{\sl Phys. Rev.} {\bf 176}, 1558 (1968).
\bibitem{Laidlaw}
	M. G. G. Laidlaw and C. M. DeWitt, 
	{\sl Phys. Rev.} {\bf D3}, 1375 (1971).
\bibitem{Mackey}
	G. W. Mackey,
	{\sl Induced representations of groups and quantum mechanics}
	(Benjamin, New York 1969).
\bibitem{OK}
	Y. Ohnuki and S. Kitakado
	{\sl J. Math. Phys.} {\bf 34}, 2827 (1993).
\bibitem{MT}
	M. S. Marinov and M. V. Terentyev,
	{\sl Fortschritte der Physik} {\bf 27}, 511 (1979).
\bibitem{Landsman-Linden}
	N. P. Landsman and N. Linden,
	{\sl Nucl. Phys.} {\bf B365}, 121 (1991).
\bibitem{TT1}
	S. Tanimura and I. Tsutsui,
	{\sl Mod. Phys. Lett.} {\bf A10}, 2607 (1995).
\bibitem{TT2}
	S. Tanimura and I. Tsutsui,
	{\sl Ann. Phys.} {\bf 258}, 137 (1997).
\bibitem{Ikemori}
	H. Ikemori, S. Kitakado, H. Otsu and T. Sato,
	{\sl Mod. Phys. Lett.} {\bf A15}, 1203 (2000).
\bibitem{Rajeev}
	S. G. Rajeev,
	{\sl Phys. Lett.} {\bf B212}, 203 (1988).
\bibitem{Mickelsson}
	J. Mickelsson,
	{\sl Phys. Lett.} {\bf B242}, 217 (1990).
\bibitem{LM1}
	M. Lavelle and D. McMullan,
	{\sl Phys. Rep.} {\bf 279}, 1 (1997).
\bibitem{LM2}
	E. Bagan, R. Horan, M. Lavelle, and D. McMullan,
	{\sl Nucl. Phys. Proc. Suppl.} {\bf 74}, 353 (1999).
\bibitem{DeWitte}
	B. S. DeWitte,
	{\sl Rev. Mod. Phys.} {\bf 29}, 377 (1957).
\bibitem{Dowker}
	J. S. Dowker,
	{\sl J. Phys.} {\bf A3}, 451 (1970).
\bibitem{statistics of shape space}
	D. G. Kendall,
	{\sl Stat. Science} {\bf 4}, 87 (1989).
\bibitem{KT}
	H. Kobayashi and I. Tsutsui,
	{\sl Nucl. Phys.} {\bf B472}, 409 (1996).
\bibitem{Landsman}
	N. P. Landsman,
	{\sl Quantization of singular systems and incomplete motions},
	e-Print Archive: gr-qc/9807069 
\end{thebibliography}
\end{document}